# LEARNING AND EVALUATION WITHOUT ACCESS TO SCHOOLS DURING COVID-19

## G. Stefansson, A.H. Jonsdottir

*University of Iceland, Science Institute (ICELAND)*

## Abstract

The tutor-web drilling system is designed for learning so there are typically no limits on the number of attempts at improving performance. This system is used at multiple schools and universities in Iceland and Kenya, mostly for mathematics and statistics. Students earn SmileyCoin, a cryptocurrency, while studying.

In Iceland the system has typically been used by students who use their own devices to solve homework assignments during the semester, accessing the Internet-based tutor-web at http://tutor-web.net. These students typically take final exams on paper at the end of the semester.

In Kenya the system is a part of a plan to enhance mathematics education using educational technology, organised by the Smiley Charity with the African Maths Initiative. This has been done by donating servers running the tutor-web to schools and tablets to students. Typically these schools do not have Internet access so the cryptocurrency can not be used.

Innovative redesign was needed during COVID-19 in spring, 2020, since universities in Iceland were not able to host in-house finals and schools in Kenya were closed so tablets could not be donated directly to students. Remote finals were held in Iceland but the implementation was largely in the hands of the instructors. In Kenya, community libraries remained open and became a place for students to come in to study.

Innovations included using the tutor-web as a remote drilling system in place of final exams in a large undergraduate course in statistics and donating tablets to libraries in Kenya. These libraries all have access to the Internet and the students have therefore been given the option to purchase the tablet using their SmileyCoin.

This paper describes these implementations and how this unintended experiment will likely affect the future development and use of the tutor-web in both countries.

## 1. INTRODUCTION

The tutor-web[1] system is designed for research[2] and learning[3]. Its drills are primarily used for learning so there are typically no limits on the number of attempts at improving performance. An important feature is that for most drills the student is shown a detailed explanation of the solution immediately after choosing an answer option. This system is used at multiple schools and universities in Iceland and Kenya, mostly for mathematics and statistics. Students earn SmileyCoin, a cryptocurrency, while studying [4].

In Iceland students tend to use their own devices. They use these to solve homework assignments during the semester by accessing the Internet-based tutor-web at http://tutor-web.net. These students typically take final exams on paper at the end of the semester.

In Kenya the system is a part of a plan to enhance mathematics education using educational technology, organised by the Smiley Charity in collaboration with the African Maths Initiative and several individuals. This has been implemented by donating servers to schools and tablets to students. Typically these schools have not had Internet access so the cryptocurrency can not be used.

Innovative redesign was needed during COVID-19 in spring, 2020, since universities in Iceland were not able to host in-house finals and schools in Kenya were closed so tablets could not be donated directly to students.

This paper describes these new implementations and how this unintended experiment will likely affect the future development and use of the tutor-web in both countries.

## 2. METHODOLOGY

### 2.1. COVID-19 and tutor-web in Iceland

Innovative redesign was needed during COVID-19 in Iceland in spring, 2020, since universities were not able to host the usual in-house finals in May. Preceding this came announcements of closures of classrooms, lectures were moved on-line as were practice sessions and mid-terms needed to be modified.

One undergraduate course in statistics was better prepared than many since the tutor-web system was already used for drills during the semester. The tutor-web was already set up in January as a homework component, where students needed to complete a certain minimal number of drills and the resulting grade would count towards their final grade.

At closure time the first mid-term had been given as a multiple-choice in-house paper exam, but it proved fairly straight forward technically to set up both the second mid-term and the final exam as tutor-web exercises. However several design choices had to me made and some of these needed extensive collaboration with the students themselves.

#### 2.1.1. Designing drill sets

The most common method to design tutor-web drills is to set up multiple-choice items in the form of a drill set outside the system and subsequently upload the entire drill set. A drill set is commonly a collection of drills on a fairly narrow topic, but can in principle also be an overview collection on several topics.

The three main approaches to designing drill sets in tutor-web are (a) handcrafting individual items (b) using random numbers to generate an entire drill set based on a single item and (c) use a generic "check the appropriate answer" header with a choice of a correct option and several distractors, where both the correct answer and distractors are chosen randomly from a reasonably large collection of possible options.

When thousands of items are needed as in the current setup, the first approach is not feasible and it was found that, given the time available, approach (c) was the most feasible option.

For some scenarios a fourth approach can be implemented as a combination of (b) and (c), given enough resources. For example, to teach students how to interpret regression output from a statistical software, option (c) would present static regression output and ask dozens of questions such as "is the estimated slope significantly different from zero" or "what percentage of the variation in the data is explained by the model". The combined approach would be to randomly generate data, run the regression and choose one of the random items, but based on the generated data. It is seen in [5] that the combined approach should be better than (c) alone, but requires considerably greater time (and dedicated programming) to set up.

#### 2.1.2. Adapting exams to COVID-19 mode

The tutor-web has several options for assigning drills, including a choice between an exam-mode and drill-mode.

In drill-mode students can continue for as long as they like, requesting new drill items until their stopping criterion is satisfied. In general, stopping criteria may be the student's own perception of a satisfactory grade, a minimal number of drill requests set by the instructor or a minimal grade set by the instructor. On the other hand, in exam-mode, a fixed set of questions is put to the student, one item at a time in sequence and the student simply answers an item and moves on to the next one. Both modes have been used in other courses.

Given the extraordinary circumstances in 2020, considerable discussions were undertaken with students in the on-line classroom on how to proceed with the mid-term. These discussions seemed to indicate that a number of students were quite agitated with any new proposals. It was therefore decided to go with the drill-mode used earlier for homework, thus minimising any changes from what had been used earlier. Design and results from the second mid-term are discussed below, but its implementation was largely uneventful and generally considered a successful response to the COVID-19 circumstances.

Remote finals were held at the University of Iceland in May of 2020, but the implementation was largely placed in the hands of the instructors with almost no warning. Given the positive experience from the second mid-term and homework, it was decided to use the tutor-web as a remote drilling system in place of final exams in the statistics course.

It should be noted that "drill mode" implies that students can continue to request new items from a drill set for as long as they like. The item grades of 0 or 1 for each answer are combined into a grade for a drills set, using a tapered weighting scheme. For this course a fixed tapering scheme model was used. A student could potentially get a complete grade after answering exactly 7 correct answers - with no errors - but with increasing numbers of incorrect answers the taper became longer with a maximum of the 30 most recent answers used in the grade.

Students were given several days to solve each of the mid-term and the final exam. After the completion of the mid-term, students were given an interim grade composed of various pieces of homework and the two mid-terms. At that stage all students were given the option to leave it at that and get a Pass/Fail course grade or alternatively take the final exam to receive a combined numerical course grade, based on a weighted average of the interim grade and the final exam.

The mid-term had four drill sets. Each was composed of 300 drill items based on a random selection of correct options and distractors with occasional "None of the Above" or "All of the Above", which could be either correct or incorrect (NOTA+/NOTA-/AOTA+/AOTA-). The underlying number of correct choices and distractors in each drill set were (20, 48), (13, 13), (41, 58), (40, 52).

For the final exam, eight drill sets with a total of 2380 drill items were constructed. Some of these sets were based on earlier homework or the mid terms, but some were completely new. The composition of the drill sets is given in Table 1. Each drill set had a single header giving an introduction to the problem, followed by "check the most appropriate box". All the items were generated using approach (c) above, by choosing a correct option and distractors from a set, possibly appending a NOTA or AOTA, which could be correct or incorrect. The total number of distractors was chosen randomly using a truncated Poisson distribution, except in the NOTA/AOTA cases, where the NOTA/AOTA option was always the fourth and last option.

*Table 1. Number of underlying options, number of generated items and frequency of correct or incorrect "All of the Above" or "None of the Above".*

| Drillset | Correct set | Distractor set | Generated items |
| --- | --- | --- | --- |
| 1 | 43 | 60 | 280 |
| 2 | 41 | 53 | 300 |
| 3 | 15 | 24 | 300 |
| 4 | 13 | 23 | 300 |
| 5 | 45 | 62 | 300 |
| 6 | 16 | 38 | 300 |
| 7 | 26 | 35 | 300 |
| 8 | 24 | 38 | 300 |
| Total | 223 | 333 | 2380 |

|      | Correct NOTA/AOTA | Incorrect NOTA/AOTA |     |
|------|-------------------|---------------------|-----|
| NOTA | 143               | 374                 | 517 |
| AOTA | 151               | 319                 | 470 |

## 2.2. COVID-19 and tutor-web in Kenya

### 2.2.1. Designing drills

The single most important learning goal of the Kenyan project is to assist secondary school students in getting into university. The most difficult hurdle in this regard is passing the national exam, known as KCSE. The tutor-web has a generic tutorial (module) for secondary school mathematics but in 2020 a new module was added, specifically targeting the KCSE. This was implemented by taking questions from an example exam and writing a computer program for each question. The program was run to obtain drill sets of 100 drill items for each question and the entire collection of drill sets was installed as a module in the tutor-web.

This corresponds to approach (c) in the preceding section.

### 2.2.2. The Smiley Charity

A non-profit organisation, the Smiley Charity (registered in Iceland as Styrktarfélagið Broskallar), runs a project to donate servers to schools and tablets to students. In previous years this has involved designing servers to run without any Internet connection and hand-delivering tablets to students to emphasise their personal ownership of the tablets. The project runs under the name Education in a Suitcase since normally the tablets and servers have been brought as hand luggage to each site.

Considerably redesign was needed during COVID-19 since schools in Kenya remained largely closed so tablets could not be donated directly to students.

### 2.2.3. Moving from schools to libraries

In Kenya, in spite of the school closures, community libraries remained open during the COVID-19 pandemic and became a place for students to come in to study. The Smiley Charity therefore arranged with several partners in Kenya to distribute tablets to libraries. This is a considerable change in strategy since the tablets no longer go directly to students from the charity. Instead, they are purchased by partners in Kenya and donated under certain conditions to partnering libraries.

Naturally there are pros and cons to each method of giving tablets directly to students or to an organisation. In the former case the recipient and end user are well defined. However in the latter case the tablet may find much more use since it is lent to more students.

### 2.2.4. Internet connections, cryptocurrency rewards and the "Library model" of donating tablets

The tutor-web has rewards in terms of grades, like other drilling systems, but it can also reward students with an electronic token, the SmileyCoin. The SMLY is a cryptocurrency, which can be redeemed by the student and used outside the tutor-web.

In previous cases, the schools and students, which the Smiley Charity has worked with, have not had wireless connections and usually not even Internet in the schools. In these cases the blockchain is not accessible and the SmileyCoin has not been an option.

The cooperating libraries all have local WiFi and access to the Internet. The students are therefore able to earn SmileyCoin rewards as they progress towards high grades in the tutor-web.

Each library can start lending tablets to students based on much fewer donated tablets than corresponding to a regular classroom of 20-30 students. Initially therefore, each library receives only 5-10 tablets.

The primary purpose of the SmileyCoin is to encourage learning. For a reward scheme to have an impact, the reward has to make the effort required worthwhile. In earlier SmileyCoin experiments, small on-line markets have been set up where the students can purchase items such as tickets to the cinema. In the library model a different option opens up, which is to allow the diligent student to purchase the tablet itself using SmileyCoin earned by studying.

The tutor-web has two student groups, the self-registered students and students who are registered as a part of a real-world classroom. Reward schemes can be set differently for the two groups, to avoid abuse of the cryptocurrency.

These considerations lead to the **library model**: The tablets are donated to a library, which has full discretion of how the tablets are lent to students. Pre-registered student accounts are set up in the tutor-web and assigned to students by the librarian. Each student can then practise for as long as they like, including towards acing the complete KCSE drill set. Acing the KCSE drill set gives a predefined reward of 1 M SMLY. On the back of each tablet is a QR-code corresponding to a payment address. If a student scans this code they get an option to pay for this tablet with 1 M SMLY. If a library gets to the stage of students purchasing their tablets, the Smiley Charity replenishes the tablet stock.

Some care is needed in tuning the rewards in this scenario. The Smiley Charity has finite funds and the rewards can be tuned at several scales so there are a fairly large number of options on how to assign rewards for performance.

### 2.2.5. Moving to libraries

A number of issues had to be solved before the move to libraries could be undertaken, including a change in the contract with the funding agency (Ministry of Foreign Affairs in Iceland), setting up agreements with libraries and arrangements with contact persons and NGOs in Kenya. Videos explaining the new approach were generated (e.g. https://bit.ly/TheLibraryModel) and the reward settings in the tutor-web were modified.

Data on the answers to each drill item are collected in a tutor-web database. Normally these are used by instructors to monitor their students but in the present setting there are no instructors. These data are only accessible for general anonymised statistical analysis.

Monitoring links were also set up, anonymous in terms of students, yet giving incentive for some competition amongst libraries.

## 3. RESULTS

## 3.1. Performance measures in statistics course in Iceland

A total of 337 students were registered in the course until the end of the semester and thus had the option of getting an in-semester grade. Ample time was given to each component of this interim grade so any students would have been able to work their way up to a passing grade by putting enough time and effort into it.

The students were given an "exit option", by leaving the course with a pass/fail based on the interim grade and not taking the final exam. A surprising number of students chose this option and in fact mostly students with very high interim grades chose to take the final.

*Table 2. Distribution of 337 students according to semester grades (0-10 interim grade). Some of these students opted to stop with a Pass/Fail based on the semester grade whereas others took the final exam.*

| Sem. grade | 0 | 1 | 2 | 3 | 4 | 5 | 6 | 7 | 8 | 9 | 10 | Total |
|---|---|---|---|---|---|---|---|---|---|---|---|---|
| **Fails** | 43 | 10 | 4 | 2 | 2 | | | | | | | **61** |
| **Passes** | | | | | | 3 | 24 | 22 | 26 | 28 | 10 | **113** |
| **Fail w Exam** | 1 | 3 | 1 | | 1 | | | | | | | **6** |
| **Pass w exam** | | | | 1 | 1 | 4 | 2 | 5 | 8 | 31 | 105 | **157** |
| **Total** | 44 | 13 | 5 | 3 | 4 | 7 | 26 | 27 | 34 | 59 | 115 | **337** |

Table 2 describes the distribution of students according to their interim (in-semester) grades. Of the 337 students registered, 61 failed the course based on the in-semester grades. Only 7 of these asked to take a final exam to try to mediate and pass the course. The most likely explanation is that the remaining 54 students had effectively stopped attending the course, as seen from the peak frequency at 0 or 1 out on the 10-scale. The 113 who achieved a passing interim grade and decided to leave the course with a "Pass" are seen to have a totally different grade distribution with a mode at 8. Of the people taking the exam only 6 failed the course due to a combined low numeric grade, all had a very low semester grade to begin with.

It is also seen that of the 163 students who decided to take the final, 136 had in-semester grades of 9 or 10. Thus by far most of the students who decided to take the final exam were simply looking to get the high numeric grade, which they knew they deserved. This turned out to be the case: 99 of these students aced the course with a clean grade of 10, 41 got 9.5 and 18 got 5-8.5.

### 3.2. Performance measures for libraries in Kenya

*3.2.1. Competition and monitoring*

A monitoring page was set up at https://libraries.tutor-web.net/ where the number of students in each library is displayed, along with their total attempts at KCSE drills as well as the overall progress in terms of acing individual drills and the entire KCSE drill collection. The web page is refreshed with new data several times per hour and serves several purposes, one of which is to monitor the progress in the libraries in order to intervene if many students do not appear to be able to use the tutor-web to their advantage.

Acing the full collection of KCSE drills is not quite trivial: The minimal effort required to ace a single drill set is to answer the first 7 drill items correctly. The KCSE collection consists of 50 drill sets, so an outstanding student would need to answer 350 questions correctly in the first attempt. A single incorrect intermediate answer has a considerable effect on the grade and implies that most students require quite a bit more than just 7 correct answers to complete a drill set.

Experience from using the same system in various schools indicates that reality tends to set in after a while and that it is in fact quite difficult for students to maintain the stamina to fully complete something as difficult as secondary school mathematics in an on-line learning environment. It must be iterated that in the tutor-web environment it is not enough to read and claim understanding. To show mastery, the student needs to demonstrate performance based on all 50 drill sets. It is similarly quite difficult for instructors to maintain the enthusiasm to keep advising students to continue with the next drill.

A second purpose behind the web page is therefore to demonstrate the aggregate performance to all the libraries and entice staff at each library to advance the use of the systems.

Libraries were also promised increased tablet donations in response to students purchasing tablets. The exact setup was that if the libraries first received a donation of 10 tablets, then they would receive another 10 tablets whenever the first student managed to purchase one tablet. After that the Smiley Charity promised to replenish the stock of tablets in the library.

## 4. CONCLUSIONS

### 4.1.1. Experience from using the tutor-web for evaluation in a statistics course

It should be noted that the statistics course also had some projects based on analysing real data using R and delivering a report and this was an important component of the course. However, most of the evaluation was tutor-web based which is of a formative assessment type: Students are allowed to learn as they are being evaluated and can constantly improve their status during the evaluation.

The only really surprising aspect of the grade analysis in this paper is the number of students who had a very high interim grade but decided to leave the course with a "Pass" rather than take the final and obtain the high grade which they could certainly have obtained given the nature of the tutor-web. Although a formal survey has not been conducted, informal questions during web sessions seemed to indicate a high level of anxiety among the students. It is therefore quite plausible that a number of very good students had non-course-related issues, which precluded them being able to take the final exam. Many students clearly had the ability and circumstances to study from home, but it is also considered known that many students did not have this. This was also seen during some web-sessions, where family obligations made participation in sessions difficult.

### 4.1.2. Experience from using the tutor-web in libraries

Prior to the start of the project several of the librarians were optimistic that their students would easily master the technology to earn their tablets. Several partners in the project also expressed belief that the SmileyCoin rewards and option to purchase the tablet would greatly incentivise the student. The promise of increased tablet donations in response to students purchasing tablets clearly motivated library staff, as the libraries had no means of procuring tablets.

Underlying the plan of replenishing tablets is the plan of donating several hundred tablets to Kenya using existing funds, along with several new options of modifying the tutor-web system and new funding mechanisms, using the library model.

One month into the project 37 students had answered at least one drill and 25 students had aced the first drill set. The student who was farthest along had aced 35 of the KCSE drill sets and was steadily working towards acing all 50 sets.

As was to be expected with new student groups in new locations, several issues were raised and solved, including problems with moving into a new drill set. Other issues included the level of initial handholding for students and librarians alike.

The overall conclusion is very promising: There was no direct contact from the program developers to Kenya in 2020. This was the first time that the tutor-web was being used without direct support from the developers. It was also the first time that tablets were purchased locally with no setup assistance from the developers.

This study demonstrates that it is quite feasible to extend tablet-based mathematics education into libraries in low-income regions, given enthusiastic librarians, high-quality drill sets with explanations and appropriate rewards.

## ACKNOWLEDGEMENTS

A large number of individuals and institutions have made this work possible. Through the years, the projects have received funding from The Icelandic Centre for Research and from several EU H2020 grants, most recently FarFish (Horizon 2020 Framework Programme Project: 727891 — FarFish). The current initiative in the Kenyan libraries is funded by the Icelandic Ministry of Foreign Affairs, including tablet purchases and corresponding logistics.

Continuous support has been provided by the University of Iceland where the course material has been developed, and by the University of Iceland Science Institute where most of the research and development has been conducted. The current version of the tutor-web was developed by Jamie Lentin at Shuttle Thread Ltd and Jamie has also looked participated in the development of the SmileyCoin wallets.

Countless students have contributed to the tutor-web and the SmileyCoin wallet, most recently Eyþór Eiríksson, who wrote the computer programs to generate the entire set of drills for the KCSE, Jóhann Haraldsson who led the TA group for the intro stats course in the middle of COVID-19.

The African Maths Initiative is the Smiley Charity's main partner and they have led the uptake of tablet-based educational technology in Kenya. The work in the western part of Kenya is led by several individuals, particularly Zachariah Mbasu, Thomas Mawora and Maxwell Fundi. The Smiley Charity has a subsidiary in Nairobi, where the library initiative is led by Kamau Mbugua in cooperation with Prof. Evelyn Njurai of Kisii University. The library partners are the Kibera Community Library (internal project leader Mary Kinyanjui), Mathare Community Library (internal project leader Billian Ojiwa) and Kongoni community library (internal project leader Elphas Ongong'o).